\documentclass{aa}  

\usepackage{natbib}
\bibpunct{(}{)}{;}{a}{}{,} 
\usepackage{hyperref}
\hypersetup{
    colorlinks=true,
    citecolor=blue,
    linkcolor=red,
    filecolor=magenta,      
    urlcolor=cyan,
    }

\usepackage{soul}
\usepackage{xcolor}
\usepackage{graphicx}
\usepackage[varg]{txfonts}
\usepackage{array}
\usepackage[normalem]{ulem}

\begin{document}

   \title{The Great Dimming of Betelgeuse: The photosphere as revealed by tomography over the past 15 years\thanks{The reduced STELLA echelle spectra and determined radial velocity are only available in electronic form at the CDS via anonymous ftp to \url{cdsarc.cds.unistra.fr} (xx-xx) or via \url{https://cdsarc.cds.unistra.fr/cgi-bin/qcat?J/A+A/}}}

    \titlerunning{Photosphere of Betelgeuse as revealed by tomography over the past 15 years}

   \author{Daniel Jadlovský
          \inst{1,2}
          \and
          Thomas Granzer\inst{3}
          \and
          Michael Weber\inst{3}
          \and
          Kateryna Kravchenko\inst{4}
          \and
          Jiří Krtička\inst{1}
          \and
          Andrea K. Dupree\inst{5}
          \and
          Andrea Chiavassa\inst{6}
          \and
          Klaus G. Strassmeier\inst{3}
          \and
          Katja Poppenhäger\inst{3}
          }

   \institute{Department of Theoretical Physics and Astrophysics, Faculty of Science, Masaryk University, Kotl\'a\v rsk\'a 2, Brno, 611 37, Czech Republic  \\
            \email{jadlovsky@mail.muni.cz}
        \and
            European Southern Observatory (ESO), Karl-Schwarzschild Str. 2, D-85748, Garching bei München, Germany
        \and
              Leibniz-Institut für Astrophysik Potsdam (AIP),
            An der Sternwarte 16, D-14482 Potsdam, Germany
        \and
            Max Planck Institute for Extraterrestrial Physics (MPE), Giessenbachstrasse 1, D-85748, Garching bei München, Germany
        \and
            Center for Astrophysics | Harvard \& Smithsonian, 60 Garden Street, MS-15, Cambridge, MA, 02138, USA        
        \and
            Université Côte d’Azur, Observatoire de la Côte d’Azur, CNRS, Lagrange, CS 34229, 06304 Nice Cedex 4, France\\}

   \date{ }

 
  \abstract
   {Betelgeuse, a red supergiant star of semi{-}regular variability, reached a historical minimum brightness in February 2020, known as the Great Dimming. Even though the brightness has returned to the values prior to the Great Dimming now, it continues to exhibit highly unusual behavior.}
   {Understanding the long{-}term atmospheric motions of Betelgeuse and its variability could be a clue to the nature of the Great Dimming and the mass{-}loss process in red supergiants. Our goal is to study long{-}term dynamics of the photosphere, including during the Great Dimming.}
   {We applied the tomographic method, which allows different layers in the stellar atmosphere to be probed in order to reconstruct depth{-}dependent velocity fields. The method is based on the construction of spectral masks by grouping spectral lines from specific optical depths. These masks are cross{-}correlated with the observed spectra to recover the velocity field inside each atmospheric layer. }
   {We obtained about 2800 spectra over the past 15 years that were observed with the STELLA robotic telescope in Tenerife. We analyzed the variability of five different layers of Betelgeuse's photosphere. We found phase shift between the layers, as well as between the variability of velocity and photometry. The time variations of the widths of the cross{-}correlation function reveal propagation of two shockwaves during the Great Dimming. For about two years after the dimming, the timescale of variability was different between the inner and outer photospheric layers. By 2022, all the layers seemingly started to follow a similar behavior as before the dimming, but pulsating with higher frequency corresponding with the first overtone.}
   {The combination of the extensive high{-}resolution spectroscopic data set with the tomographic method revealed the variable velocity fields in the photosphere of Betelgeuse, for the first time in such detail. We were also able to find new insights related to the Great Dimming event and its aftermath, namely the discovery of another shockwave and the subsequent rearrangement of the photosphere. Our results demonstrate that powerful shocks are the triggering mechanism for episodic mass{-}loss events, which may be the missing component to explain the mass{-}loss process in red supergiants.} 

   \keywords{stars: supergiants -- stars: atmospheres -- stars: mass-loss -- shock waves -- techniques: spectroscopic }
   
   \maketitle
%

\section{Introduction}

Betelgeuse ($ \alpha$ Orionis) is a nearby red supergiant (RSG) star. Its semi{-}regular variability is characterized by two main timescales common to many RSGs. These are (i) a shorter period of roughly 400 days, attributed to the radial pulsations of the atmosphere in the fundamental mode \citep[FM,][]{kiss06,chatys19,joyce20,jadlovsky22}; and (ii) a longer period of about 2100 days ($ 5.6 \: \rm  years $), which is of a less certain origin. The longer period has been attributed to the flow timescales of giant convective cells \citep{stothers10}, which \citet{lopez_ariste22} found some evidence for through spectropolarimetry imaging. It has also been attributed to processes that drive a similar mode of variability in asymptotic giant branch (AGB) stars, where it is known as the long secondary period (LSP), most likely caused by non{}radial, low{-}degree oscillations due to gravity modes. However, other mechanisms were also proposed: for example, magnetic activity \citep{wood00, wood04}. In many RSGs, \citet{kiss06} and \citet{chatys19} also found higher overtones of the FM period, primarily the first overtone. For Betelgeuse, \citet{joyce20}  detected a photometric period of about 185 days, which is likely the first overtone mode. \citet{granzer22} found a period of about 216 days based on radial velocities as well as evidence for the second overtone. However, for all the modes of variability, the observed length and amplitude of each cycle is not constant \citep{kiss06}.

Betelgeuse underwent an unprecedented historical minimum of brightness, the so{-}called Great Dimming, that culminated in February 2020 when its brightness dropped by $ V \sim 1.6 \: \rm mag$ \citep{guinan20}. Interferometric observations by \citet{montarg21} revealed that the star dimmed asymmetrically, as the southern hemisphere of Betelgeuse was much darker than its northern counterpart. Many theories have been put forward to explain the Great Dimming, such as a significant decrease in temperature or the formation of star spots \citep{dharma20, harper20}, an increased molecular opacity in the outer layers \citep{kravchenko21}, or a surface mass-loss event \citep{dupree20,dupree22}. Interferometric observations appear to be compatible with both leading theories: the cold spot on the surface as well as a dust clump above the photosphere \citep{montarg21,cannon23,drevon24}.

In the detailed study that combined the previous findings, \citet{dupree22} suggested that the Great Dimming was caused by a shockwave (reported by \citet{kravchenko21}) that formed in the photosphere and resulted in substantial surface mass ejection of matter along our line of sight in September{-}November, 2019. The ejected material reached the outer atmosphere in about six months, eventually creating the molecules and dust as it reached cooler regions. \citet{dupree22} and \citet{jadlovsky22} also observed that following the Great Dimming, the star pulsated with shorter periods than 400 days. 

New models by \citet{davies21}, which successfully incorporated stellar wind and, for the first time, reproduced extended atmospheres and related observed spectral features of RSGs, generally support the scenario proposed by \citet{dupree22}, although the authors show that the dimming can be explained by TiO molecule absorption, without the need for dust condensation in the ejected material. Nonetheless, \citet{taniguchi22} reported an increased extinction by oxygen{-}rich dust that contributed to the Great Dimming.

In the present paper, we intend to reveal long{-}term atmospheric motions of Betelgeuse and its variability by applying the tomographic method by \citet{alvarez01} and \citet{kravchenko18} to a time{ }series of high{-}resolution spectra of Betelgeuse. In Sect. \ref{chapter:observations}, we describe our data set and the application of the tomographic method. In Sect. \ref{chapter:res}, we present the results. In Sect. \ref{chapter:dimming}, we take a more detailed look at the Great Dimming and its aftermath, based on our data. In Sect. \ref{chapter:discussion}, we discuss the shocks and excitation of higher overtones, and we give our conclusions in Sect. \ref{chapter:conclusions}.


\section{Observations and methodology}
\label{chapter:observations}
\subsection{Observations}

We employed spectra observed with the Stellar Activity (STELLA) echelle spectrograph (SES) mounted on two fully robotic 1.2\,m telescopes at the Izanã Observatory in Tenerife, which is operated by AIP \citep{stella,stella_1}. The resolving power is $R\approx$55\,000 with a three{-}pixel sampling per resolution element. The spectra typically have signal{-}to{-}noise ratios (S/N) of $100-600$. Exposure times range from five to 20 seconds. The spectra cover the $390{-}880 \: \rm nm$ wavelength range. As of January, 2024, the full time series of Betelgeuse consists of about 2800 spectra (about 6000 exposures in total) obtained during the years 2008{--}2024. Most of these observations were originally presented by \citet{granzer21,granzer22}, where the telescope, observations, and methodology are also briefly described. 

\subsection{Methodology}
\subsubsection{Envelope tomography}
\label{chapter:tomography}
The tomographic method was first introduced by \citet{alvarez00,alvarez01}. The method allows us to probe different layers in the stellar atmosphere and recover the corresponding velocity fields. This is accomplished by sorting spectral lines into groups (masks) based on their formation depth. The masks are cross{-}correlated with observed or synthetic stellar spectra and provide velocity fields within different atmospheric layers.

Cross{-}correlation (CCF) functions of some Mira{-}type stars appear double{-}peaked during the maximum brightness \citep{alvarez00}. One of the main goals of using the tomographic method was to show that this feature can be explained by the passing of shockwaves through the atmosphere, which is called the Schwarzschild scenario \citep{schwarz52}; that is, the second peak would correspond to the rising shock front. \citet{alvarez00, alvarez01, alvarez01b} successfully applied the method and recovered the Schwarzschild scenario on spatial and temporal scales in several Mira{-}type stars. The only limitation of the method was that it assumed that the line depression originates in layers where the optical depth $ \tau_\lambda$ is $2/3 $. However, that was shown not to be the case for weak lines \citep{albrow96}. It was also difficult to make quantitative predictions as the geometric radius associated with each mask was not known.

\citet{kravchenko18} improved the method by including the computation of the contribution function to calculate the formation depth of spectral lines. The method was then validated on a 3D radiative{-}hydrodynamics simulation of an RSG atmosphere \citep{chiavassa11}. \citet{kravchenko19} applied the tomographic method to high{-}resolution spectroscopic time{-}series observations of the RSG star $ \mu $ Cep to interpret its photometric variability. Finally, \citet{kravchenko20} validated the tomographic method on spectro{-}interferometric VLTI/AMBER observations by recovering a link between optical and geometrical depth scales based on comparison to dynamical model atmospheres.

Afterward, \citet{kravchenko21} also applied the tomographic method to high{-}resolution spectroscopic time{-}series observations of Betelgeuse taken with the HERMES spectrograph from 2015{-}2020 to interpret its photometric variability, including the Great Dimming event. The resolution of HERMES is 86\,000, which is larger than resolution of STELLA, and their data set included about 30 spectra. They found two shockwaves preceding the dimming (in February, 2018 and January, 2019), the latter amplifying the former. They proposed that the succession of these two shocks resulted in an outflow (as shown by velocity gradients) and an increase of molecular opacities as their interpretation of the dimming. They did not rule out the dust scenario, but they could not confirm it based on spectroscopic observations. 

To apply the tomographic method in this paper, we used a spectral template\footnote{Downloaded from \url{https://marcs.oreme.org}} (spherical symmetry, $ T_{ \rm eff} = 3400 \: \rm K$, $ \log{g} = -0.4$, $ v_{\rm microturb} = 2 \: \rm km \, s ^{-1}$ and solar metallicity) calculated with the Turbospectrum synthesis code \citep{plez12} with MARCS models \citep{gustaf08} and VALD3 atomic and molecular line lists \citep{ryab15}. Furthermore, to create tomographic masks that correspond to different layers of Betelgeuse's photosphere, we used the same masks as \citet{kravchenko19}. Therefore, the spectral template for Betelgeuse was modified to include only the lines from each tomographic mask; that is, we created five new templates with selected lines. This was achieved by removing all lines from the template that were not included in the masks (for each mask). This is done differently than in \citet{kravchenko19, kravchenko21}, where Dirac distribution was used as line profiles.

Table \ref{table:masks} lists the masks used in this paper. We used the same masks as \citet{kravchenko21}, except for mask C1, where we had to use fewer lines in order to increase the signal. Lines in mask C1 have the highest optical depths, while each following mask has lower values of optical depths than the one before it, all relative to a reference optical depth $\tau_{0}$ computed at $5000  \, \rm \mathring{A}$, as described in \citet{kravchenko18}. Therefore, mask C1 corresponds to the innermost layer of the probed photosphere and mask C5 corresponds to the outermost layer.

\begin{table}[t]
        \caption{Lists of spectral lines from \citet{kravchenko19} used in each mask, $\tau_{0}$ limits, and numbers of spectra, which were left from the original data set after successful cross{-}correlations and subsequent cleaning.} 
        \label{table:masks} 
        \setlength{\extrarowheight}{3pt}
        \begin{tabular}{cccc}
\hline
\text{Mask} & \text{$\log{\tau_{0}}$ limits }  & \text{N. of lines }  & \text{N. of spectra} 
\\
\hline
C1 & $-1.0 < \log{\tau_{0}} \leq 0.0 $ & 135 & 1679 \\ 
C2 & $-2.0 < \log{\tau_{0}} \leq -1.0$ & 1750 & 2290 \\
C3 & $-3.0 < \log{\tau_{0}} \leq -2.0 $ & 1199 & 2309 \\
C4 & $-4.0 < \log{\tau_{0}} \leq -3.0 $ & 433 & 2311 \\
C5 & $-4.6 < \log{\tau_{0}} \leq -4.0 $ & 378 & 2312 \\
        \hline
        \end{tabular}
\end{table}

\subsubsection{Cross{-}correlation functions}
\label{chaper:obs}

The data were reduced using the IRAF{-}based pipeline SESDR in its version 4.0 \citep{stella_3,stella_2}. All spectra were corrected for echelle blaze and wavelength calibrated with consecutively recorded Th{-}Ar spectra. To derive cross{-}correlation functions (CCFs), the radial velocity determination is done order{-}by{-}order, that is, each echelle order is cross{-}correlated with the mask separately. Afterward, the cross{-}correlation functions are summed up, and the radial velocity is determined by a Gaussian fit (for a single observation). 

Many of the spectra were removed due to failed guiding of the telescope or other various instrument-related issues (about $7\% $). The numbers of successfully cross{-}correlated spectra left after the cleaning are listed in Table \ref{table:masks}. 

\section{Results}
\label{chapter:res}

This section is focused on characterizing the overall variability of Betelgeuse based on the application of the tomographic method to our data set. Our results are also compared to those of \citet{kravchenko21}, who applied the tomographic method to Betelgeuse for the first time.

\begin{figure*}[htbp]
    \includegraphics[width=1.0\textwidth, keepaspectratio]{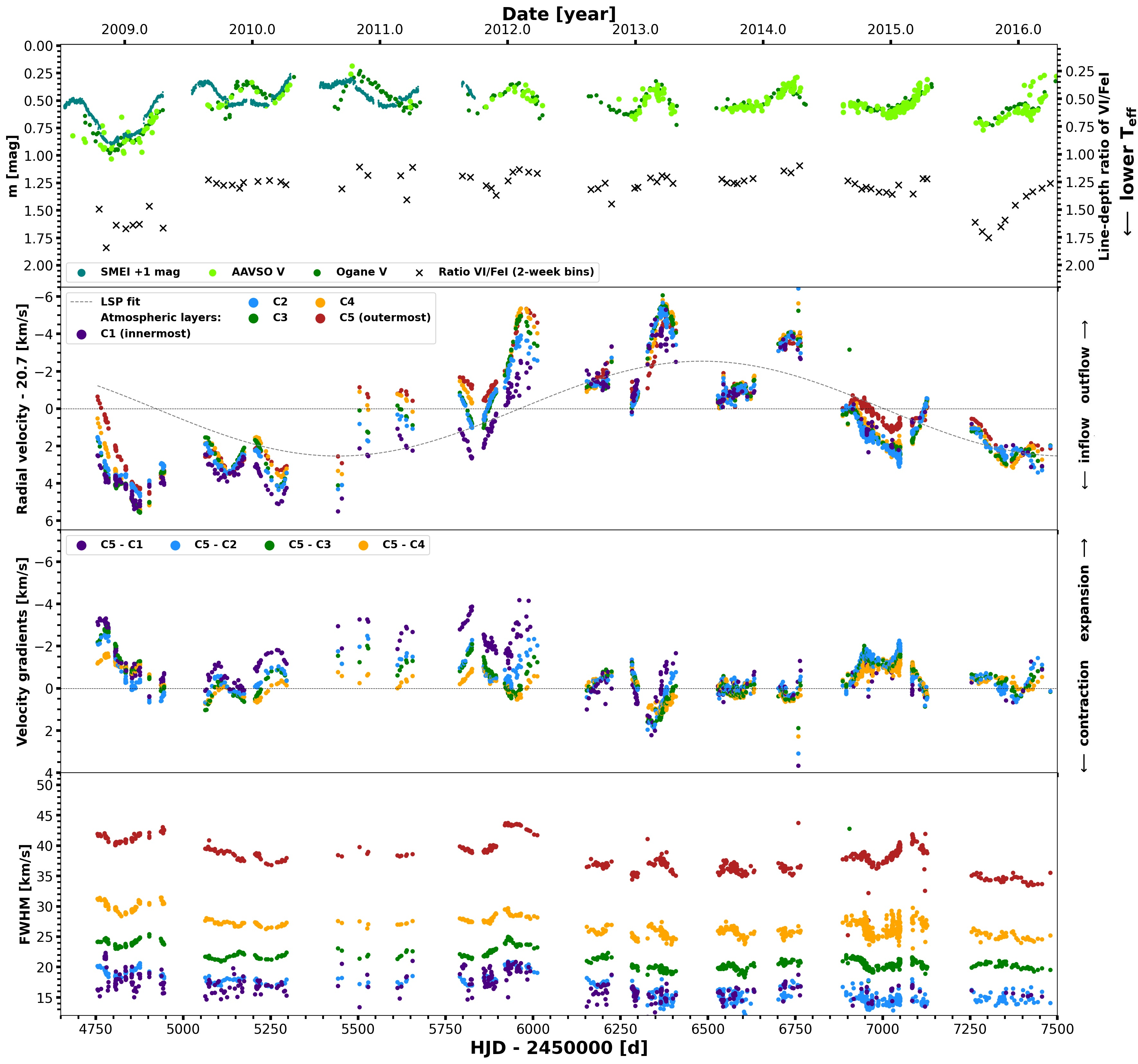}
    \caption{Variability of Betelgeuse from 2008{--}2016
    {\em First panel:}
    Light curve comprising of data from AAVSO, Ogane, and SMEI from 2008{-}2015. Temperature is also included via the line{-}depth ratio of \ion{V}{i} and \ion{Fe}{i}.
    {\em Second panel:}
    Radial velocity plot determined from STELLA spectra. The spectra were cross{-}correlated with five masks corresponding to five different layers of photosphere. The same color-coding is applied in the fourth panel. The velocity of the center{ }of{ }mass ($ 20.7 \: \rm km \, s ^{-1} $) was subtracted. Both main periods, the FM and LSP, are clearly evident in the data, while we can also see the presence of higher overtones of the FM period between years 2009 and 2010.
    {\em Third panel:}
    Gradients of radial velocity from masks in second panel. At most times the gradients are not very high.
    {\em Fourth panel:}
    FWHMs of CCFs. The FWHMs are correlated with the variability of the star, to a certain extent. }

    \label{fig:mag_rv_1}
\end{figure*}

\subsection{Brightness}
The first panels of Figs. \ref{fig:mag_rv_1} and \ref{fig:mag_rv_2} show the evolution of brightness adopted from the American Association of Variable Star Observers (AAVSO), Solar Mass Ejection Imager \citep[SMEI,][]{smei04, hick07}, and Ogane Hikari Observatory \citep{ogane22}. The first two data sets consist of more data than plotted in the figures. For SMEI, a cleaning algorithm as described in \citet{paunz21} and \citet{jadlovsky22} was applied. 

The combination of these data sets covers the semi{}regular photometric variability of the star and the Great Dimming well. Both main modes of variability are clearly visible in the data, as well as variability of a shorter timescale than the FM period.

\subsection{Relative temperature changes}
The line{-}depth ratio of \ion{V}{i} ($6251.83  \, \rm \mathring{A}$) and \ion{Fe}{i} ($6252.57 \, \rm \mathring{A}$) is sensitive to temperature changes due to excitation potentials of corresponding levels; therefore, it can be used as a proxy for temperature. The ratio becomes larger for decreasing temperature in the photosphere \citep{gray00,gray08}. Calibrating these ratios to absolute values of effective temperature would be difficult. However, the relative variability of temperature is sufficient for us, as we primarily use it for plotting hysteresis loops in Sect. \ref{chapter:loops}.

The ratios are plotted alongside the brightness in the upper panel of Figs. \ref{fig:mag_rv_1} and \ref{fig:mag_rv_2}. The derived relative temperature changes are mostly in phase with the photometric variability, similarly to the derived effective temperature in previous studies \citep[e.g.,][]{kravchenko19, dupree22}.

\subsection{Radial velocities}
\begin{figure*}[htbp]
    \includegraphics[width=1.0\textwidth, keepaspectratio]{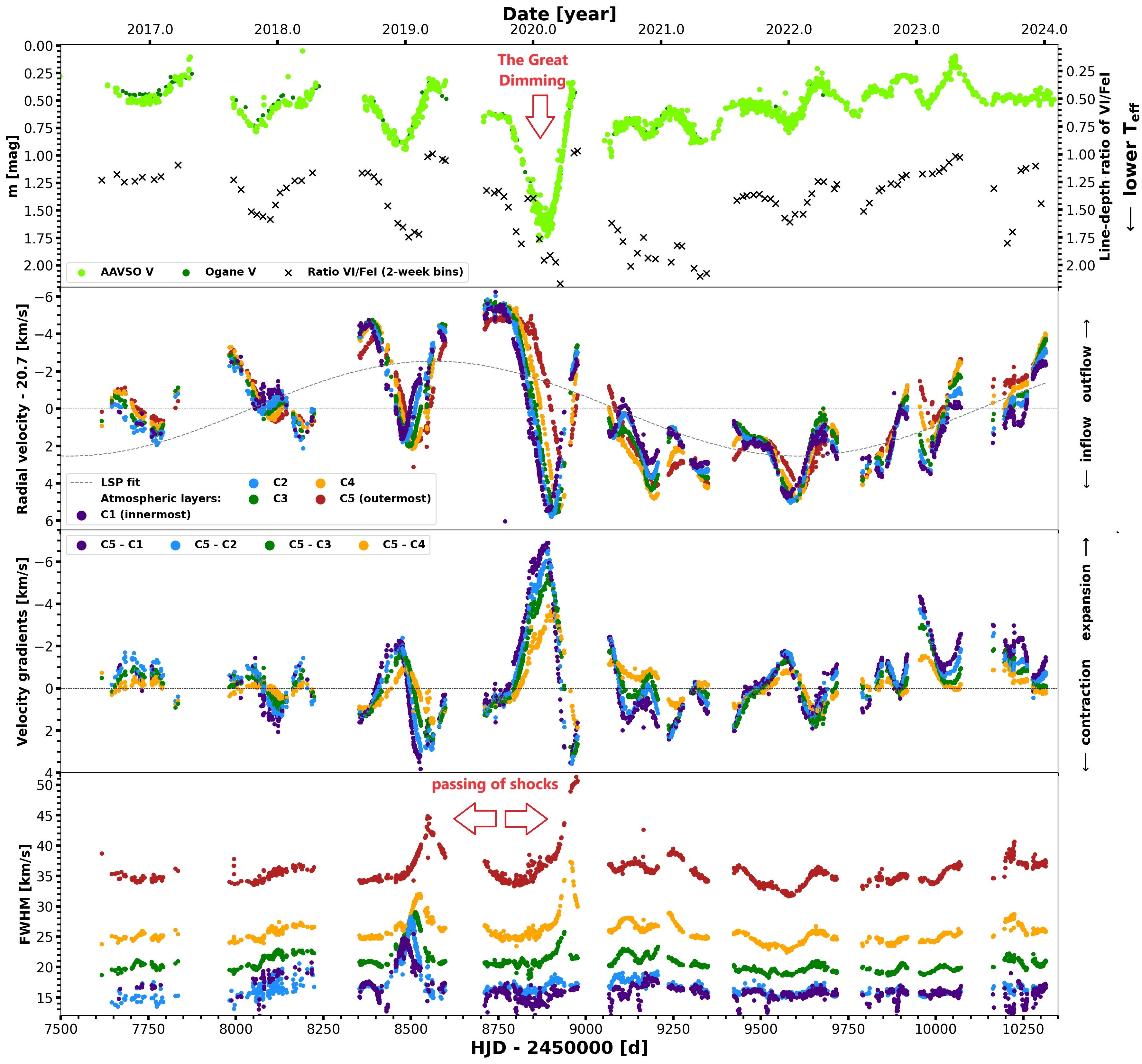}
    \caption{Same as Fig. \ref{fig:mag_rv_1}, but for 2016{-}2024.
    {\em First panel:} Light curve shows historical Great Dimming.
    {\em Second panel:} Prior to Dimming, we can see a major long{-}lasting outflow velocity, followed by extreme infall velocity.
    {\em Third panel:} During Great Dimming, gradients were uniquely negative, suggesting a strong expansion of the inner layers.
    {\em Fourth panel:}
    Before and after Dimming, there are two significant peaks in the FWHMs, while the latter peak was much stronger. Both peaks are caused by a passing shockwave.
    }

    \label{fig:mag_rv_2}
\end{figure*}

The second panels of Figs. \ref{fig:mag_rv_1} and \ref{fig:mag_rv_2} reveal kinematics of five different layers of the photosphere using five cross{-}correlation tomographic masks from Table \ref{table:masks}. Thanks to the abundance of observations, we are able to see the long{-}term evolution of different layers of Betelgeuse in such a detail for the first time. The zero{-}point is set at $ 20.7 \: \rm km \, s ^{-1}; $ this corresponds to the center{-}of{-}mass (CoM) velocity of Betelgeuse (the average value between \citet{harper17} and \citet{kervella18}).

The usual maximal peak{-}to{-}peak amplitudes of velocity variations in all photospheric layers reach up to $ \sim 5 \: \rm km \, s ^{-1} $, but often the amplitudes are smaller, especially when there is a variability within shorter timescales than the FM period. All these modes of variability are also modulated by the LSP period, which is clearly evident in the data and is represented in Figs. \ref{fig:mag_rv_1} and \ref{fig:mag_rv_2} by a sinusoidal curve. This curve was determined based on a Lomb{-}Scargle periodogram performed on the radial velocity data \citep{granzer22}. Parameters of the sinusoidal curve are a period of $ 2081 \: \rm days$, an amplitude of $ 2.54 \: \rm km \, s ^{-1}, $ and initial phase of $ 0.71 $. The shorter periods of variability, likely the higher overtones of the FM period, appear to become more prominent when the photospheric velocities are lower than the velocity of CoM of the star for several cycles of the FM period. This is shown in Fig. \ref{fig:mag_rv_1} from 2008{-}2010 and in Fig. \ref{fig:mag_rv_2} from 2020{-}2023, and it is further discussed in Sects. \ref{chapter:dimming} and \ref{chapter:discussion} .

When rising material starts to propagate through the photosphere, the innermost layer, C1, is the first to be affected, and it is mostly in phase with brightness $V$ and temperature variability; namely, the minima occur in approximately the same point of time. The time shift of velocity minima between the innermost (and thus also the brightness and temperature minima) and outermost layers can be up to about $ 50 \: \rm d $ ($ 10 \: \rm d $ between neighboring layers), which is quite common from 2017{-}2023, and the overall scenario is well illustrated by the minimum at the beginning of 2019. The values of time shifts correspond to the ones reported by \citet{kravchenko21}. However, from 2008{-}2017, the time shifts are much smaller, only up to about $ 10 \: \rm d $ in total, while the velocity shifts between the layers are also less significant. Nonetheless, we have fewer minima covered by the data in this time period. During the velocity maxima, the time and amplitude shifts are also shorter. Because the (observed) length and amplitude of each cycle is not constant, we only give approximate values of time and amplitude shifts between individual radial velocity curves.

\subsection{Velocity gradients}
The third panels of Figs. \ref{fig:mag_rv_1} and \ref{fig:mag_rv_2} show curves of velocity gradients, that is, the evolution of velocities relative to the outermost layer C5. The gradients were calculated in the same manner as in \citet{kravchenko21}; that is, we defined gradients for layers C1{-}C4 as $ v_{\rm C5} - v_{\rm C_{i}} $, where $\rm i = 1, 2, 3, 4$. Thus, when the gradients are negative ($ v_{\rm C5} - v_{\rm C_{i}} < 0 $), they signify the expansion of regions over which the gradient was taken, whereas when the gradients are positive ($ v_{\rm C5} - v_{\rm C_{i}}  > 0 $), they signify contraction.

The gradients are quite small at most times, within $ 2 \: \rm km \, s ^{-1} $ of the zero point, and quite often even at the zero point. In other words, at most times all the layers are moving with nearly the same velocity, such as between the years 2014 and 2017. 

In cycles when the velocity variability is stronger, the gradients reveal to us that when the material is rising, the inner layers of the photosphere are typically being contracted (see, e.g., the positive gradients at the beginning of years 2013 and 2019 in Figs. \ref{fig:mag_rv_1} and \ref{fig:mag_rv_2}, respectively). On the other hand, when the material in the layers is infalling, all the layers usually move with a similar velocity. Except during the Great Dimming, we only rarely see a significant expansion of the atmospheric layers.

\subsection{FWHMs}
\begin{figure*}[htbp]
    \includegraphics[width=1\textwidth, keepaspectratio]{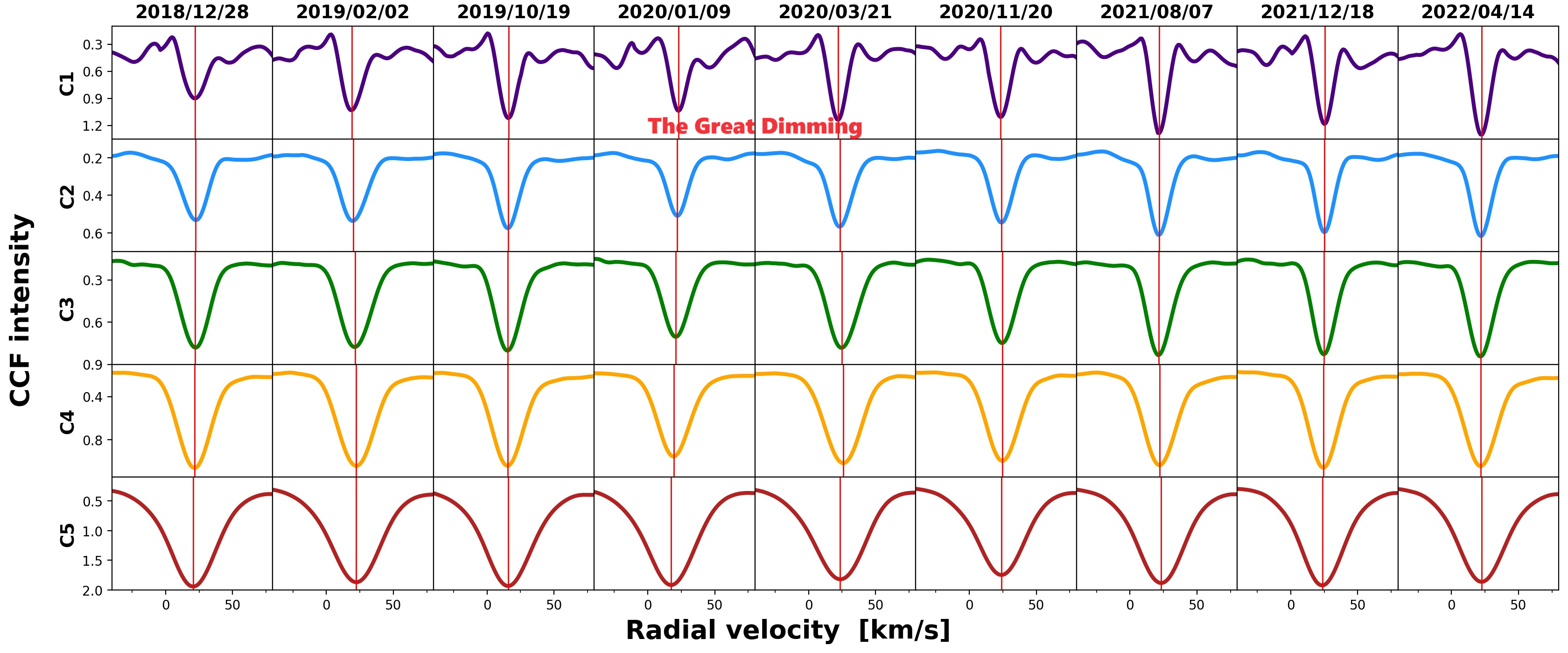}
    \caption{Preview of CCF profiles sampled during cycles related to Dimming. The values of CCF intensity (y{-}axis) are inverted. Vertical red lines show radial velocity determined. In CCFs of mask C1, we can often clearly see that the central components are asymmetric, suggesting a secondary component. Other inner layers sometimes also show weak asymmetries.  } 
    \label{fig:ccf_preview}
\end{figure*}

The bottom panel of Figs. \ref{fig:mag_rv_1} and \ref{fig:mag_rv_2} shows the full width at half maximum (FWHM) of the CCFs. The average FWHM in layer C1 is about $ 17 \: \rm km \, s ^{-1} $. The FWHM increases in the upper layers, up to about $ 37 \: \rm km \, s ^{-1} $ in layer C5. Our CCF profiles resemble the ones by \citet{kravchenko21}, which reported secondary components shifted by $ 10-15 \: \rm km \, s ^{-1} $ with respect to the main peak. We have also found many similar weak asymmetries, especially in CCFs of mask 1. A sample of our CCFs showing the typical asymmetries is plotted in Fig. \ref{fig:ccf_preview}.

Similarly to \citet{kravchenko21}, we find the FWHM of CCF to be a reliable indicator of a passing shockwave. We can see a significant increase (by a factor of about 1.3) in FWHMs at the beginning of 2019 (Fig. \ref{fig:mag_rv_2}, also shown in \citet{kravchenko21}), followed by an even more unprecedented increase (by a factor of about 1.5 in outer layers) at the beginning of 2020. Meanwhile, in the rest of the data, the variability of the FWHMs is considerably smaller. This makes it clear that the Great Dimming was truly an extraordinary and unprecedented event. While the first shockwave in the beginning of 2019 was already reported by \citet{kravchenko21}, the stronger shockwave in 2020 has not yet been reported in any study. 

For both shockwaves, we can see a time shift of the emergence of a shockwave in a similar manner to that seen in the RVs; that is, the shockwave is moving from the innermost mask to the uppermost mask. However, there are some differences between the shockwaves in 2019 and 2020. During the propagation of the first shockwave, the FWHM peak of the innermost layer C1 arises in a similar point of time as its velocity minimum, and the peaks of outer masks are increasingly more time shifted in relation to layer C1. In total, the time shift between layers C1 and C5 is about $ 75 \: \rm d $. Likewise, we can see a similar time shift during the shockwave in 2020, even though the FWHM peaks are larger in outer layers compared to the previous shockwave, while in the innermost layers (C1 and C2) the peaks are barely visible at all, and in layer C3 the peak is still smaller than during the previous shock. This suggests that the initial low{-}amplitude shock steepened significantly as it propagated through the photosphere. The shock was amplified by a factor of 1.5 by the time the shock front reached photospheric layers with a reference optical depth $ \log{\tau_{0}}$ smaller than $-3.0$.

The shockwave in February 2018 reported by \citet{kravchenko21} does not seem to be represented in our data by any significant FWHM peak (at least not one comparable to the two main peaks), although there is definitely a small increase of FWHM in 2018. That suggests that this shockwave was of a much smaller amplitude than the two that accompanied the dimming. Supposing this was indeed a shock of a smaller scale, then similar shocks of weaker amplitudes are quite frequent, as in Figs. \ref{fig:mag_rv_1} and \ref{fig:mag_rv_2} we can see many peaks at this variability scale. 

Additionally, we also remark that resolution of HERMES is about 86\,000, which is much higher than that of STELLA. Therefore, it is expected that our instrument would be less sensitive to asymmetries and possible line-doubling events. 

\subsection{Hysteresis loops}
\label{chapter:loops}
The phase shift between variations of temperature and velocity results in hysteresis loops, that is, the evolution of velocity and temperature during a (photometric) cycle. \citet{gray08} was able to construct hysteresis loops based on observations of Betelgeuse and proposed the following interpretation. During a regular pulsation cycle, the material starts heating (for the beginning of a cycle defined in the last temperature minimum), while the velocity is constant. Eventually, once the temperature is high enough, the material starts to rise up in the atmosphere. When the material is rising, it begins to cool, and hence it starts to infall again after some time. However, this interpretation assumes a stationary convection, and it thus cannot explain the observable time{-}dependent effects: specifically the phase shift between the temperature and velocity variations, as demonstrated by \citet{kravchenko19} and \citet{kravchenko21}. Based on simulations and theoretical predictions, \citet{kravchenko19} investigated the role of convection and acoustic waves in driving the hysteresis loops. They showed that the timescales of hysteresis loops are similar to the sound{-}crossing timescales in outer layers. Therefore, they proposed that hysteresis loops have an acoustic origin. All these papers also showed that the loops turn counterclockwise and that timescales of loops are similar to the the FM period.

We constructed hysteresis loops for Betelgeuse, and it appears that the behavior described in the aforementioned papers can generally be observed in many of the cycles, even though the exact shapes and amplitudes of each loop are often quite different from each other. The results are available in Appendices \ref{fig:loops}, \ref{fig:loops_2}, and \ref{fig:loops_3}. Cases when a clear determination of the shape or length of a loop was not possible originate in the lack of data (typically in older cycles) and poor data sampling. Additionally, the determination of the onset of cycles (based on photometric minima) is not very precise due to the complex variability of the star. 

Notwithstanding all these issues, it is quite apparent from the data that the oscillations shorter than the FM period rarely form full hysteresis loops. Usually, such curves seem to cover about half of a loop. On the other hand, the FM cycles seem to form full loops. Therefore, this suggests that the timescales of hysteresis loops are similar to the those of the FM period.

Furthermore, in Appendix \ref{fig:loops_merged} we show the evolution of velocity and temperature for the LSP cycles. This is done by binning the velocity and relative temperature changes in 400{ }d intervals, which approximately corresponds to an average length of the FM period. Quite surprisingly, the material in the photosphere also appears to form hysteresis loops during this timescale, superimposed over the shorter loops, although it is not certain whether the first LSP cycle forms a full loop. These longer loops could correspond to the turnover of material in giant convection cells on the surface.

\section{Great Dimming}
\label{chapter:dimming}

\subsection{Emergence}

\begin{figure*}[htbp]
    \includegraphics[width=1\textwidth, keepaspectratio]{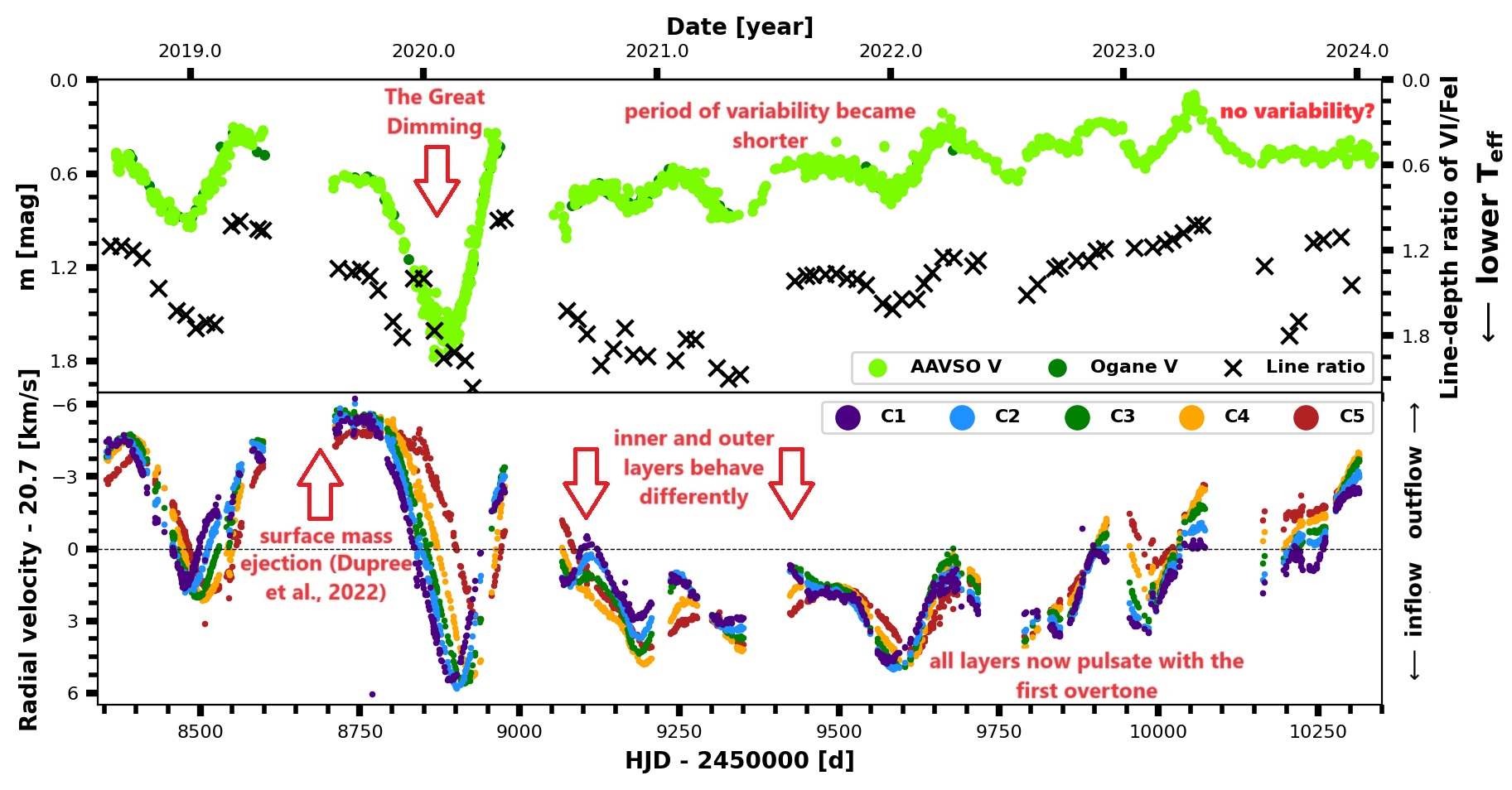}
    \caption{Detailed view of Great Dimming and its aftermath from Fig. \ref{fig:mag_rv_2}. The upper panel shows variability of brightness and temperature; the lower panel shows variability of velocity. The first red arrow denotes the unusual outflow that lasted $ \sim $ 200 d. During this outflow, the ejection of material as proposed by \cite{dupree22} could easily happen. What followed were extreme infall velocities, while the maximum infall velocities of inner layers were higher than those of upper layers, which is unprecedented in the data. In the aftermath of the dimming and the second shockwave, the outer and inner layers behaved differently. Inner layers (C1{-}C3) exhibited a separation of local outflow maxima corresponding to the first overtone (the third and fourth red arrows), whereas the outer layers appeared to have a substantially longer separation, likely corresponding to the FM. Since the beginning of 2022, the layers appear more synchronized with each other again, but the dominant mode of pulsations is now the first overtone. }
    \label{fig:dimming_aftermath}
\end{figure*}

We revealed several unique features that appeared during the Great Dimming (a closer look in Fig. \ref{fig:dimming_aftermath}). Firstly, there was the powerful shockwave \citep{kravchenko21} that first appeared in the innermost layer, C1, at the beginning of 2019 and then continued to propagate upward. Simultaneously, it caused a strong outflow of rising material (or enhanced it) that reached its maximum by the middle of 2019, with a peak{-}to{-}peak amplitude of about $ 7 \: \rm km \, s ^{-1} $ in some masks. Although the outflow velocity during the peak was quite similar to some of the preceding maxima, what made this event special was that the outflow velocity stayed at near maximum values for $ \sim 200 \: \rm d $ (the first red arrow in Fig. \ref{fig:dimming_aftermath}). This is when material of higher density would be ejected from the atmosphere as reported by \citet{dupree20,dupree22}. \citet{granzer22} reported that the pulsations have already been excited to higher overtones during this time.

What followed was a massive change of velocity; the velocity of layers C1{-}C4 changed quite abruptly by about $ 10 \: \rm km \, s ^{-1} $ between October, 2019 and April, 2020, while for layer C5 it was by about $ 8 \: \rm km \, s ^{-1} $. Most importantly, the inner layers were infalling much faster than the layer C5, causing the inner layers to expand dramatically (as shown by the extreme gradients). When the maximum infall velocity was reached by April, 2020, the inner layers, especially C2{-}C4, abnormally (compared to other cycles) had greater infall velocity than layer C5. 

In summary, the sequence of events that led to the Great Dimming began at the beginning of 2019 with the appearance of the shockwave in the photosphere \citep{kravchenko21}, followed by an enhanced chromospheric flux in September-November 2019 \citep{dupree22}. The brightness reached its historical minimum in February 2020 \citep{guinan20}, followed by a minima of photospheric velocities in April, 2020 \citep{granzer22}, and then also by a maximum outflow velocity of stellar wind at the end of March \citep{jadlovsky22}. Meanwhile, the brightness of Betelgeuse quickly began to restore its normal values. Regardless, the events that followed the dimming also continued to be unprecedented.

\subsection{Aftermath}
By the time the innermost layers reached maximum infall velocity during the Great Dimming, a rather small peak appeared in the FWHMs of C1{-}C2, possibly as the origin of the next shockwave, which eventually became even stronger than the first one (based on FWHMs behavior). By the time mask C5 reached its maximum infall velocity, the shockwave already appeared to be passing through mask C3, but the height of the FWHMs peak was still smaller than in the same mask during the previous shockwave. Simultaneously, the layers were starting to contract again. By the time the velocities of all layers were rising again, the shockwave passed through layers C4 and C5 in April-May 2020. The passing shockwave appears to reach its maximum FWHM in layer C5 in the same time the data coverage ends. In these two layers, the shockwave was much stronger than the previous one, and the strongest in the entire data set. 

\citet{dupree22} reported that the atmosphere was left less dense following the surface mass ejection. That could help explain the appearance of the new shock directly after the Great Dimming, and especially its amplified strength. However, the results by \citet{dupree22} show that unlike during the previous event, this shock was not followed by a comparable enhancement of chromospheric flux. There was only a slight enhancement of the flux in August, 2020.

Either way, the resulting outflow seems to have one of the highest peak{-}to{-}peak amplitudes in the data. The amplitude of the peak cannot be determined precisely, as the peak happened some time between May{ and }August, 2020, a time period not covered by our observations. Fortunately, the photometric data by \citet{taniguchi22} reveal that the maximum brightness was reached at approximately the moment our data coverage ends, just after the beginning of May, 2020. Thus, the maximum outflow velocity was probably reached by the middle of the observation gap, which suggests that by that time, the rising material was able to reach quite high values of outflow velocity. It seems this outflow velocity maximum followed about $ 250 \: \rm d $ after the previous one, which is shorter than the FM period. 

The second shock led to a remarkable divergence of the inner (C1{-}C3) and outer (C4{-}C5) layers of the atmosphere, which is shown in detail in Fig. \ref{fig:dimming_aftermath}. The outer layers continued their previous movement, while the inner layers were oscillating with shorter modes of variability of $100-200 \: \rm d $, likely the first or other low{-}order overtones. This unprecedented dissonance continued for about a year and half. By 2022, all the layers of the photosphere were finally synchronized, in the same way as before the dimming. However, this time the shorter mode of pulsations of about $ 200$ d was dominant \citep{dupree22, jadlovsky22,granzer22}, which is very likely the first overtone. 

However, at the beginning of 2023, the gradients show that inner layers were being significantly expanded (the second strongest expansion after the dimming). This suggests that another major event might have occurred. In April, 2023, this was followed by another brightening. Furthermore, over the last six months (since July, 2023), the brightness surprisingly shows no significant variations, while the temperature and radial velocity continue its variability.

The LSP of about $ 2100 $ d appears to remain active, but whether the FM period regains its dominance remains to be seen. Based on the previous behavior of the star, it seems likely that the FM shall indeed return. Hydrodynamic simulations by \citet{macleod23} suggest that the first overtone will eventually be reduced by the convective damping in about five years (post-dimming).

\section{Discussion}
\label{chapter:discussion}

\subsection{Shockwaves}
The spectral lines considered here come from higher up in the photosphere where the reference optical depth is smaller than one. In these layers of RSGs, the photospheric activity is characterized by interactions between non{}radial pulsations and shockwaves \citep{chiavassa10}. Based on 3D radiative{-}hydrodynamics simulations of RSG stars \citep{chiavassa11,kravchenko19,chiavassa24} and AGB stars \citep{frey17, lilje17,chiavassa24}, the shockwaves originate from acoustic waves produced in stellar interior by convection. As the wave hits the surface, it is slowed down and compressed due to the drop in temperature and sound speed. It increases its amplitude due to the decrease in density and thus turns into a shock, propagating to the outer atmosphere. The density is lower in outer layers, which causes the shockwaves to become even stronger. Consequently, the influence on the photosphere would also become more significant. As we can see in the bottom panel of Fig. \ref{fig:mag_rv_2}, both major shockwaves indeed become stronger as they reach the upper layers.

Mira{-}type stars follow the Schwarzschild scenario (as described in Sect. \ref{chapter:tomography}); that is, evident line doubling occurs in the CCFs, revealing the upward motion of the shock front, when a shockwave is passing through an atmosphere \citep{alvarez00}. However, as shown by \citet{josselin07} and \citet{kravchenko19}, the CCF profiles of RSGs do not clearly follow the Schwarzschild scenario. Instead, the secondary components primarily cause CCFs to become asymmetric, as in these stars the motions within the atmosphere are more complex. The asymmetries are related to supersonic velocity fields due to the motion of giant convective cells and possibly also non{-}spherically symmetric shockwaves. Based on 3D radiative{-}hydrodynamics simulations, \citet{kravchenko19} demonstrated that the intensity of secondary components is directly related to the emitting surface on the star. For example, when a significant fraction of the surface is rising, a blue secondary component appears in a CCF. Additionally, \citet{ma23} demonstrated that large{-}scale convective motions can produce components that may be mistaken for the signs of rotation.

Similarly to \citet{kravchenko21}, we can see in our CCFs that even during the dimming event there is no obvious hint of the Schwarzschild scenario in Betelgeuse, and rather than visible line{-}doubling events we just observe weak asymmetries (which may be due to resolution). Nonetheless, the FWHMs demonstrate that two powerful shockwaves appeared in Betelgeuse before and after the dimming, while the first one is likely responsible for the dimming itself.

\subsection{Excitation and damping of higher overtones}
\begin{figure}[t]
    \includegraphics[width=0.5\textwidth, keepaspectratio]{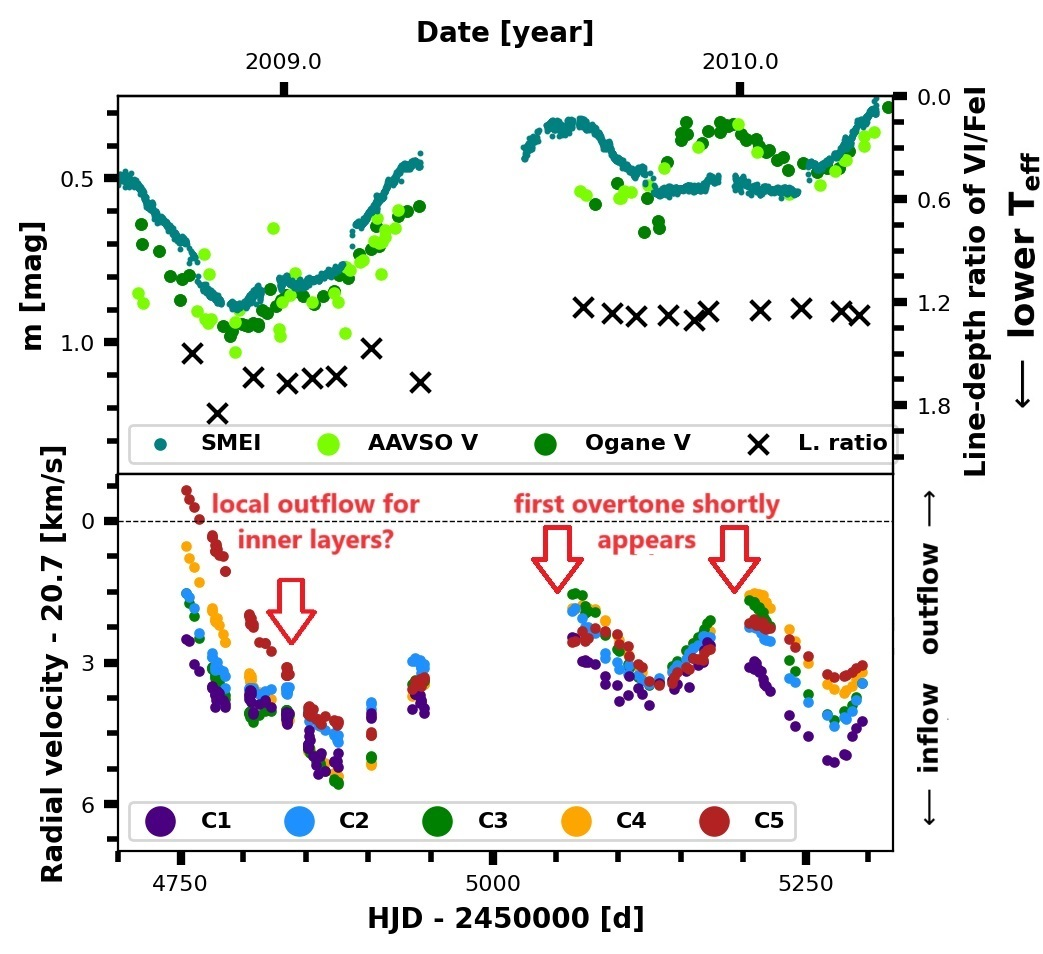}
    \caption{Detail from Fig. \ref{fig:mag_rv_1} showing presence of low{-}order overtones long before the dimming. The upper panel shows variability of brightness and temperature; lower panel shows variability of velocity. The second and third red arrows denote the estimated times of maxima, which are delayed only by about 150 days from each other. The first red arrow denotes a possible local maximum outflow velocity of inner layers (C1{-}C3), while the two outermost layers appear not to have a local maximum of velocity. Thus, the behavior of the atmospheric layers slightly resembles the aftermath of dimming. If we had more data, we could see for how long the low{-}order overtones remained prominent.}
    \label{fig:overtone}
\end{figure}

As we show in Sect. \ref{chapter:res}, the higher overtones of the FM period are not observed only after the dimming, but occasionally even before, usually when the photospheric layers are long{-}term infalling, as in Fig. \ref{fig:overtone}. Furthermore, it appears that full hysteresis loops are not formed when higher overtones dominate.
This perhaps suggests that when the material is long{-}term infalling, different processes from beneath the photosphere are capable of exciting the motion of photospheric material to higher overtones. As \citet{joyce20} proposed, a non{}linear excitation may occur; that is, the layers would eventually accumulate enough energy and momentum to compress the material beneath the photosphere. The compressed material would heat and thus impact convective structure near the surface, altering the energy flow beneath the surface. 

Assuming that the strength of the two shocks during the dimming is not unique, the shocks could be the source of additional energy or heat that dissipates into the atmosphere. As \citet{granzer22} showed, the mode of pulsations began to change shortly before the dimming, at a similar time to when the first powerful shockwave emerges in our data (Fig. \ref{fig:mag_rv_2}), while after the second shockwave the change of pulsations became even more apparent. The shocks should generally appear during a rising part of the light curve \citep{kravchenko19, kravchenko21}. The first half of the data (Fig. \ref{fig:mag_rv_1}) does not show such a powerful shockwave, although it is not impossible that such a shock may have occurred in a time gap in the data, or before the STELLA data collection began. 

\section{Conclusions}
\label{chapter:conclusions}
Thanks to the enormous data set of spectral observations by STELLA{-}SES combined with the tomographic method, we were able to analyze the long{-}term evolution of different layers of Betelgeuse's photosphere with unprecedented detail. We revealed many insights into the variability of the RSG photospheres, pulsations, formation of shockwaves, its passing through the photosphere, and a correlation of different layers of photosphere with the brightness variability and properties of the CCFs. 

We determined the radial velocity of five different layers of Betelgeuse's photosphere over the past 15 years using the tomographic method. The phase shift of radial velocities between the neighboring photospheric layers can reach about ten days (therefore 50 days in total), although during a regular cycle the phase shifts are smaller. The variability of the innermost photospheric layer C1 is approximately in phase with the photometric variations (in filter $V$); therefore, the phase shifts relative to the brightness are the same. The shock{-}crossing timescale in the photosphere appears to be longer, up to 75 days, until the shock reaches the outermost layer. The determined relative temperature variability based on a line{-}depth ratio of \ion{V}{i} and \ion{Fe}{i} \citep{gray00, gray08} is mostly in phase with photometric variations. 

Similarly to \citet{granzer22}, we find that the photospheric velocity is indeed variable on two primary timescales - the LSP of about $ 2100 \: \rm d $ and FM of about $ 400 \: \rm d $- while during some cycles the low{-}order overtones become considerably more significant, especially after the dimming. The low{-}order overtones seem to be more prominent when the photospheric layers are long{-}term infalling relative to the CoM of Betelgeuse. This possibly suggests that the mode of pulsations may be excited to higher overtones by processes that take place deeper in the atmosphere.

The timescales of hysteresis loops, which relate the phase shift between variations of velocity and temperature, are similar to the timescale of the FM period, and thus we are able to support that the photometric variations are linked to acoustic waves in a similar manner to that  proposed by \citet{kravchenko19} for $\mu $ Cep. Meanwhile, the oscillations of higher overtones appear to fail to form complete loops, which may be a clue as to why the higher overtones are being damped. Curiously, the LSP also appears to form a hysteresis loop. This could be used as evidence that the LSP period is indeed due to the motion of material in giant convection cells.

We revealed two powerful shockwaves related to the Great Dimming. The first one (reported by \citet{kravchenko21}) is likely the major progenitor of the dimming, causing a major long{-}lasting outflow of material \citep{granzer22,dupree22}, while the second shockwave emerged after the dimming and contributed to the changes of the atmosphere. The outflow caused by the first shockwave was followed by an extreme infall velocity of all layers. By the time of maximum infall velocity, a second shockwave emerged. This shock was even more powerful than the first one, at least in the outer layers. The second shockwave lead to another significant outflow. These events resulted in major changes to the pulsation characteristics of the photosphere. However, these changes did not take effect simultaneously in all the layers. For about a year and half, the pulsation mode of inner layers was already excited to the first overtone, or to some other low{-}order overtone, while outer layers remained less affected, apparently continuing their previous movement. The dissonance ended by the beginning of 2022, when all the layers now clearly pulsated with the first overtone. Therefore, the rearrangement of the photosphere appeared complete. Based on the previous behavior of the star, we expect the FM to eventually regain its dominance. Nonetheless, the behavior of the star remains unprecedented, as this was followed by the second strongest expansion of the inner layers (as shown by the gradients) in the beginning of 2023 and a significant brightening of the star in April 2023. Finally, for the last six months (since July, 2023), the star has not exhibited any significant variations of brightness.

The CCF profiles show weak asymmetries rather than clear line doubling, likely related to passing shockwaves or convective motions. Similarly to \citet{kravchenko19}, we used the FWHMs of CCFs for the detection of shocks in the photoshere. The increased FWHMs revealed two powerful shockwaves during the dimming, which are clearly unprecedented relative to the rest of the data. The FWHMs also show smaller scale variability, which could be related to shocks of lower amplitudes or convective motions.

Our results demonstrate that the triggering mechanism for mass{-}loss events are powerful shocks, while the higher overtones of radial pulsations may be excited in the process. The episodic gaseous outflows are likely the missing component to explain the mass{-}loss process in RSGs \citep{humphreys22}. 

This study further validates the potential of the tomographic method. Most importantly, it reveals many new major insights into the nature of RSGs. We intend to follow up on these results in a forthcoming work where we will compare the results to 3D radiative{-}hydrodynamics simulations.

\begin{acknowledgements}
     We thank the editor and anonymous referee for useful comments that improved the quality of the paper.

     Based on data obtained with the STELLA robotic telescopes in Tenerife, an AIP facility jointly operated by AIP and IAC.

     We acknowledge with thanks the variable star observations from the AAVSO International Database contributed by observers worldwide and used in this research.

     DJ acknowledges support from the ERASMUS+ programme of the European Union and from ESO-MEYS Training Programme of the Czech Republic.

     KP acknowledges funding from the German \textit{Leibniz Community} under project P67/2018.

     AC acknowledges support from the French National Research Agency (ANR) funded project PEPPER (ANR-20-CE31-0002)
\end{acknowledgements}

%
%
\bibliographystyle{aa} 
\bibliography{bibliography} 

%

\appendix

\begin{figure*}[htbp]
\section{Hysteresis loops} 

    \includegraphics[width=1\textwidth, keepaspectratio]{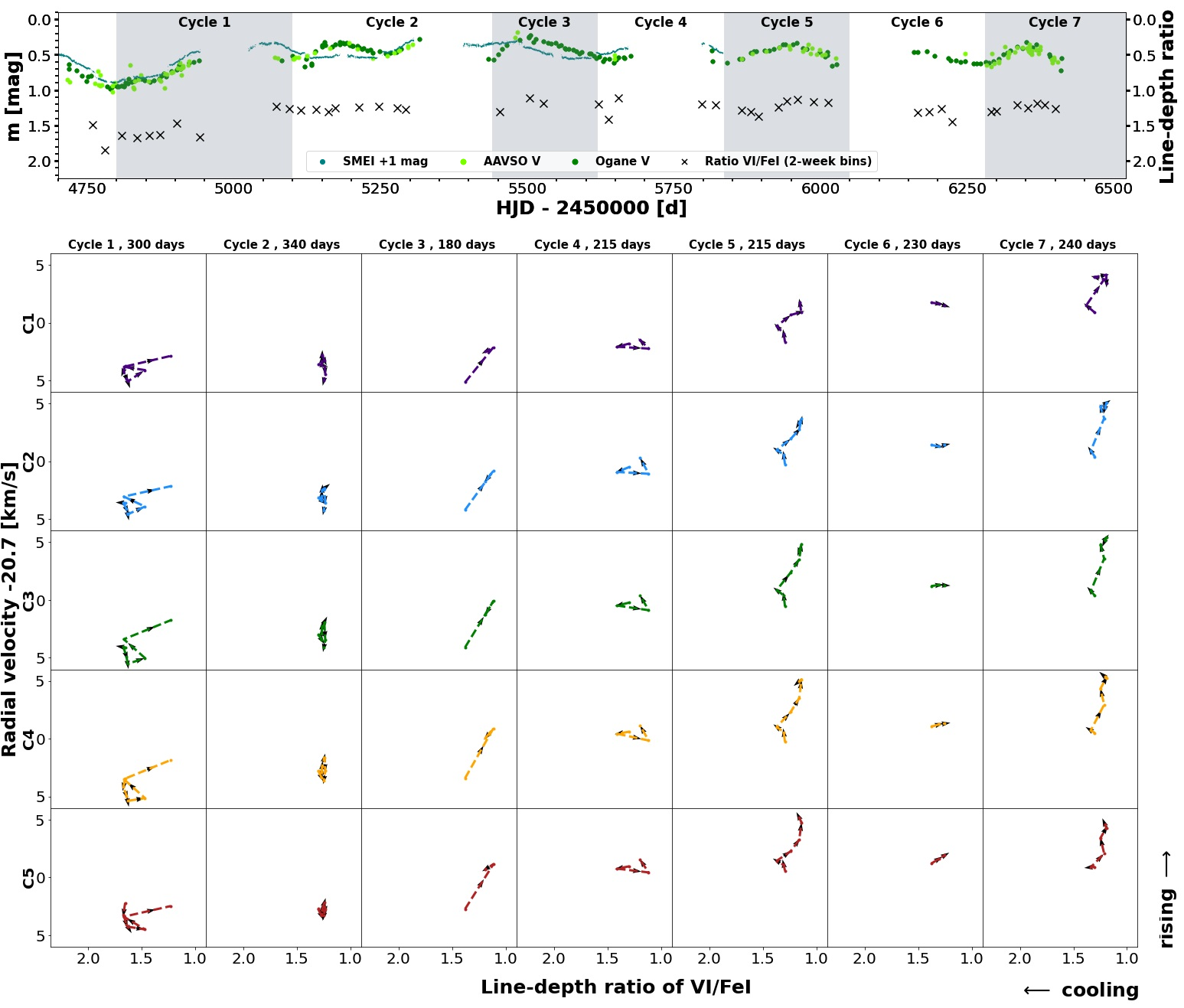}
    \caption{Hysteresis loops for photometric cycles.
    {\em Upper panel:}
    Estimated photometric cycles along with the line-depth ratios of \ion{V}{i} and \ion{Fe}{i}. The ratios are determined from the spectra. Consequently, the ratios also show which part of a photometric cycle is covered by the data.
    {\em Lower panels:}
    Hysteresis loops relate the phase shift between the variations of velocity and temperature. To demonstrate the evolution of material in atmospheric layers more clearly, the values were binned in two{-}week intervals. Most cycles lack significant parts of the loops due to an insufficient number of measurements (especially in the first half of the data) and observation gaps; therefore, in many cases full loops are not observed. The black arrows indicate the direction of a loop, which almost always is as we would expect, that is, counter{}clockwise. Continued in Fig. \ref{fig:loops_2}.}
    \label{fig:loops}
\end{figure*}

\begin{figure*}[htbp]

    \includegraphics[width=1\textwidth, keepaspectratio]{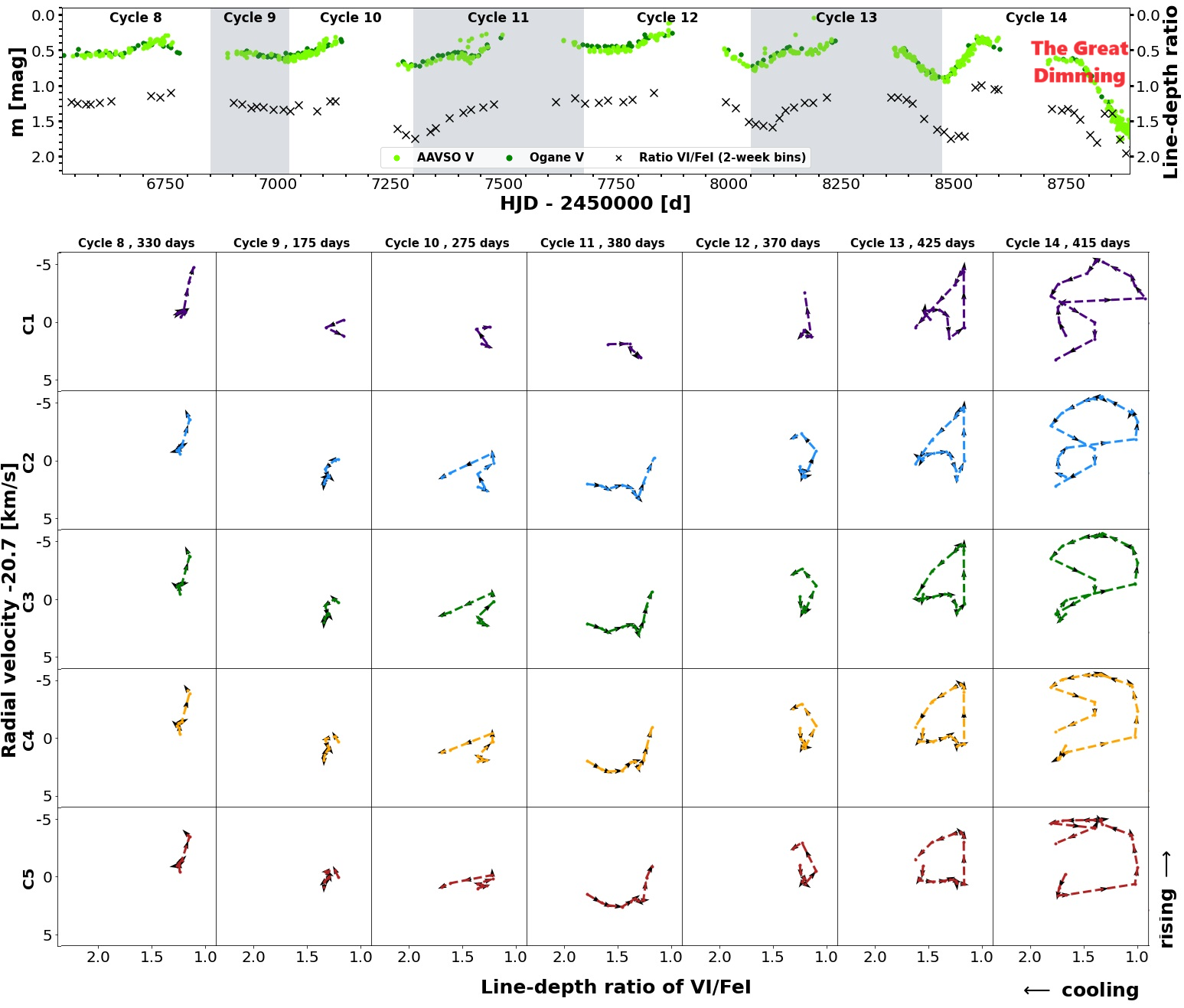}
    \caption{Hysteresis loops for photometric cycles; continued from Fig. \ref{fig:loops}.  }
    \label{fig:loops_2}
\end{figure*}

\begin{figure*}[htbp]

    \includegraphics[width=1\textwidth, keepaspectratio]{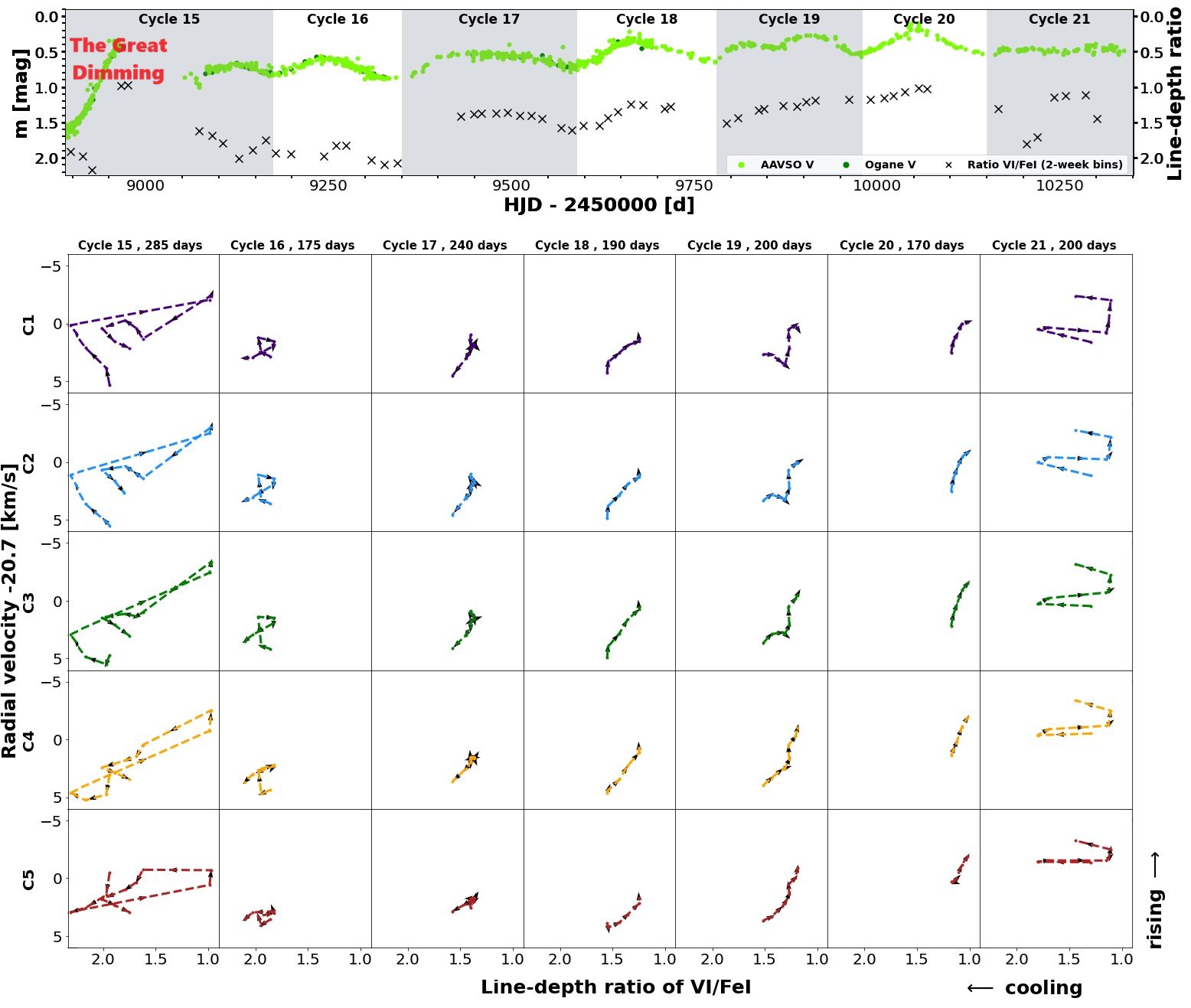}
    \caption{Hysteresis loops for photometric cycles; continued from Fig. \ref{fig:loops_2}.  }
    \label{fig:loops_3}
\end{figure*}

\begin{figure}[htbp]

    \includegraphics[width=0.5\textwidth, keepaspectratio]{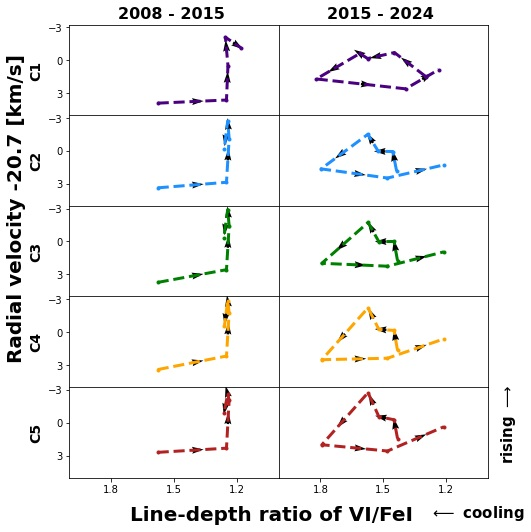}
    \caption{Phase shift between variations of velocity and temperature forms hysteresis loops. The loops are shown for approximate LSP cycles and plotted for different layers of the photosphere. The values were binned in 400 d intervals. The shape of the loop of the first cycle (left panels) is affected by data sampling. Nonetheless, the first cycle also suggests that material is returning to its starting position (except in layer C1). }
    \label{fig:loops_merged}
\end{figure}

\end{document}